# Magnetic Dipole Moment of Neutrino

Samina S. Masood
Department of Physics, University of Houston Clear Lake, Houston TX 77058
masood@uhcl.edu

**ABSTRACT**

We recalculate the magnetic moment of neutrinos in a hot and dense medium. The magnetic dipole moment of neutrinos is modified at high temperature and chemical potential. We show that the magnetic dipole moment of electron neutrino does not get a significant contribution from thermal background to meet the cosmological bound. However, chemical potential contribution to the magnetic moment is non-ignorable even when chemical potential is an order of magnitude greater than the electron mass. It is demonstrated that this effect is more significant in the models with an extended Higgs sector through neutrino mixing.

## 1. Introduction

Neutrinos were originally discovered as massless neutral fermions. These fermions can only interact weakly and conserve individual lepton number in the standard electroweak model (SEWM). Neutrinos are massless left-handed fermions in SEWM and are called Weyl neutrinos. These left-handed neutrinos can interact weakly with leptons (of the same flavor) and quarks. On the other hand, the neutrino oscillation (flavor or spin) is only allowed with nonzero mass of neutrino, though this mass can be small enough to give ignorable contributions to beta decay. However, neutrinos can oscillate between different flavors if they have nonzero mass and a right-handed partner. A very tiny nonzero mass of neutrino leads to the numerous extensions of the standard model and opens up new venues of research in high energy physics with its tremendous applications to astroparticle physics and cosmology. Minimally extended standard model (MESM) still obeys the law of conservation of individual lepton number. In this paper we look at the MESM with nonzero tiny Dirac mass which does not change the SEWM other than adding an inert right-handed neutrino as a singlet in each generation.

Massless neutrinos are represented by Weyl spinners with one degree of freedom. These Weyl neutrinos have to strictly obey the conservation of individual lepton number. However, nonzero mass [1] of neutrino was originally suggested to resolve Solar Neutrino Problem (SNP). Neutrino mixing or neutrino oscillation is a natural consequence of massive neutrino, though extremely small. SEWM allows two types of massive neutrinos; Dirac type of neutrinos bring in a right-handed neutrino as an anti-particle with equal and opposite mass; whereas Majorana mass adds a heavy right-handed neutrino for seesaw mechanism. Massive neutrino, with a Dirac mass can establish a higher order coupling with the magnetic field through radiative corrections and has nonzero magnetic moment which is induced by the corresponding charged lepton. The Majorana neutrino can have the transition magnetic moment that is also associated with the coupling of neutrino with charged leptons and is called transition magnetic moment. However, the transition magnetic moment can be expressed in terms of the magnetic dipole moment of Dirac neutrino [2-6].

Superkamiokande claimed the experimental evidence of neutrino oscillations [7-9] indicating nonzero mass of neutrino as a possible solution of SNP. This nonzero mass of neutrino encouraged phenomenologists to study the properties of massive neutrinos in detail and look for other applications of nonzero mass of neutrino. If the neutrino mass is not exactly equal to zero then we can think about the existence of right-handed neutrino as well. Now, to accommodate this right handed neutrino in the SM, we have to look for different extensions of the SM. These tiny masses of neutrinos, though they are not expected to be high enough to change the physical results significantly provide enough motivations to look at the extensions of the standard electroweak model, even more seriously. In this paper, we will work on MESM, mainly. All the Feynman diagrams and the corresponding rules of the SM remain unchanged in this model.

MESM can further be extended with increased particle sectors associated with the modified conservation rules. In one of the simplest model, we have to extend the Higgs sector to accommodate the neutrino doublet in the theory at the expense of the individual lepton number. Total lepton number is always conserved. This new charged Higgs can be added as a singlet [10] or as a doublet [11]. We would refer to the last one as the Higgs doublet model (HDM) throughout this paper and look at it as an example of possible extension of SEWM, with embedded neutrino mixing. There are several other extensions of the standard model which we do not include in our discussions because we mainly want to calculate the quantum statistical background effects on the properties of massive neutrinos. Although the neutrino masses are small enough to be ignored in most cases, we still have to investigate the behavior of massive neutrinos in astrophysical and cosmological environments where they behave differently with this tiny mass. Astronomical environments can be extremely hot and dense with strong magnetic field. We study the behavior of Weyl neutrino in the constant magnetic field. Our basic scheme of work is real-time formalism [12-18] where we deal with the real particles processes in the heat bath of real particles at moderate energies.

This paper has been organized in a way that the next section is devoted to the study of the properties of massless neutrino in a strong magnetic field in extremely dense systems. Properties of Dirac neutrinos at finite temperature and density are calculated in section 3 for different ranges of temperature and chemical potential. Majorana neutrinos are not discussed in detail because the transition magnetic moment of Majorana neutrinos [2] can be expressed in terms of the magnetic moment of Dirac neutrino. In the last sections we discuss all of the calculated results and their implications.

## 2. Weyl Neutrino

Weyl Neutrinos are massless neutrinos and conserve the individual lepton number. It does not interact with the medium directly. Weyl neutrinos, being massless neutral particles, do not couple with the magnetic field, even at extremely high temperatures. However, they can acquire a dynamically generated shielding mass like other neutral particles [15-18]due to its interaction with electrons in an extremely dense rotating system with a high concentration of charged particles such as electrons. With this effective mass, a neutrino will behave as a Dirac particle with the effective mass [19]

$$m_{\text{eff}} = \frac{g^2 e|B|\mu_e}{(2\pi)^2 (m_W^2 - e|B|)} = \frac{g^2 \mu_e eB/m_W^2}{(2\pi)^2 (1 - eB/m_W^2)} \tag{1}$$

$B$ is the constant magnetic field and $\mu_e$ gives the chemical potential of electrons in the background. Eq. (1) can later be inserted in the magnetic moment expression to get the magnetic moment of neutrino, corresponding to the dynamically generated mass and is treated as Dirac mass .This effective mass of neutrino increases with increasing magnetic field, whereas $W$ boson mass sets an upper limit on the magnetic field, that is; $e|B| < m_W^2$ to get the physically acceptable effective mass. In this equation, Eq.(1) is used to calculate the rate of change of this mass with the constant magnetic field comes out to be [19]

$$\frac{dm_{\text{eff}}}{dB} = \frac{g^2 e\mu_e}{(2\pi)^2 (m_W^2 - eB)} \left(1 + \frac{eB}{(m_W^2 - eB)}\right), \tag{2}$$

Eq. (2) gives the coupling of neutrino through its dynamically generated mass $m_{\text{eff}}$ and $\frac{dm_{\text{eff}}}{dB}$ which is proportional to the magnetic moment of neutrino and is proportional to the chemical potential of the corresponding charged lepton of the same generation. Rate of change of the effective mass with respect to the constant magnetic field of the medium is independent of temperature. Weyl neutrino cannot interact with thermal medium. For $m_e < |B| \ll m_W^2$, we can write it as

$$\frac{dm_{\text{eff}}}{dB} = \frac{g^2 e\mu_e}{(2\pi)^2 m_W^2} \left(1 + \frac{eB}{m_W^2}\right) \approx \frac{g^2 e\mu_e}{(2\pi)^2 m_W^2}, \tag{3}$$

and, for very large magnetic field

$$\frac{dm_{\text{eff}}}{dB} = \frac{g^2 e^2 \mu_e B}{(2\pi)^2 (m_W^2 - eB)^2}, \tag{4}$$

which is simply ignorable until the chemical potential $\mu_e$ is very large. Eq.(2) can also be

rewritten as

$$\frac{dm_{eff}}{dB} = \frac{g^2 e\mu_e \left(eB/m_W^2\right)}{(2\pi)^2 m_W^2 \left(1 - eB/m_W^2\right)^2}.\qquad(5)$$

The above equation indicates that the coupling with the weak magnetic field depends on the magnetic field itself. However, for strong field, the effective magnetic moment is constant and will be significant only if the chemical potential is large enough.

Just to understand the functional behavior of the effective mass, we plot the effective mass (Figure 1) and the rate of change of the effective mass (Figure 2) with respect to magnetic field by setting (for the constant chemical potential)

$$\frac{g^2 e\mu_e}{(2\pi)^2 m_W^2} \approx 1 \qquad(6)$$

The plot of Eq. (1) indicates that the effective mass of neutrino becomes significant only for an extremely large magnetic field (Millions of Tesla) which is only possible for extremely large chemical potential of electrons. Just to look at the magnetic field dependence, we choose constant density of electrons associated with the tremendously large chemical potential. Stable systems usually cannot acquire such high values of density and magnetic fields, however, these conditions cannot be ruled out during the process of generation of superdense stars.

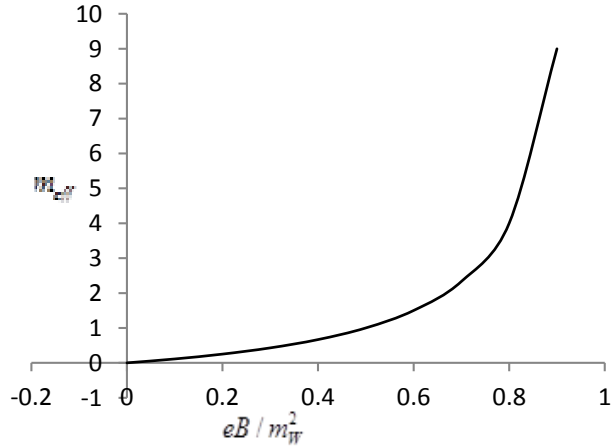

*Figure 1: Rate of change of the effective mass with respect to the magnetic field.*

It can be clearly seen from the graph that the magnetic field effect is only significant for extremely large magnetic field which is only possible for unusually dense systems.

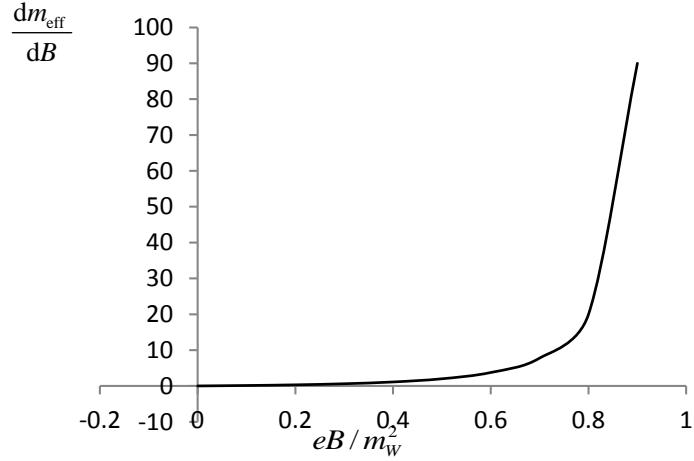

*Figure.2: Plot of the rate pf change of the effective mass with respect to the magnetic field as a function of the magnetic field.*

### 3. Dirac Neutrino

Induced magnetic dipole moment is associated with neutral Dirac particle through its interaction with the corresponding charged lepton. In the MESM, the lepton flavor mixing is not allowed just as the SEWM. Right-handed neutrino is the only additional singlet in this model and is inert in nature. So all the Feynman rules are obeyed and basic diagrams of the SEWM are studied. Dirac neutrino has a nonzero magnetic moment in vacuum [4]. The boson propagator in the real-time formalism can be written as [12-18]

$$D_B^\beta = \frac{1}{k^2 + i\varepsilon} - 2\pi i \delta(k^2) n_B(k) \qquad (7a)$$

and the fermion propagator as,

$$S_F^\beta = (p - m_\ell)\left[\frac{1}{p^2 - m_\ell^2 + i\varepsilon} + 2\pi i \delta(p^2 - m_\ell^2) n_F(p)\right] \qquad (7b)$$

with the Bose-Einstein distribution,

$$n_B(k) = \frac{1}{e^{\beta k_0} - 1} \qquad (8a)$$

for massless vector bosons. This automatically leads to the vanishing mass density and hence the chemical potential. The Fermi-Dirac distribution is given by [20]

$$n_F(p) = \frac{\theta(p_0)}{e^{\beta(p_0 - \mu)} + 1} + \frac{\theta(-p_0)}{e^{\beta(p_0 + \mu)} + 1} \qquad (8b)$$

such that the chemical potential of fermions is always equal and opposite to the corresponding anti-fermions in CP symmetric background. However, Eq. (8b) can still work (for comparable values of particle-antiparticle concentration) even if CP symmetry breaks. Since all the particle species are in thermal equilibrium, it is convenient to expand the distribution functions in powers of $m\beta$, where m is the mass of the corresponding particles and $\beta = T^{-1}$.µ, T and m correspond to the properties of electrons and are expressed in the same units. We obtain [14-15]

$$n_F \xrightarrow{\mu \prec T} \sum_{n=1}^{\infty}(-1)^n e^{n\beta(\mu-p_0)} \qquad (9a)$$

and in the limit of $\mu \succ T$ [16,17]

$$n_F \xrightarrow{\mu > T} \theta(\mu - p_0) + \sum_{n=1}^{\infty}(-1)^n e^{-n\beta(\mu-p_0)} \qquad (9b)$$

for particles. Distribution functions for antiparticles, in the same approximations give

$$n_F \xrightarrow{\mu \prec T} \sum_{n=1}^{\infty}(-1)^n e^{-n\beta(\mu+p_0)} \qquad (10a)$$

and

$$n_F \xrightarrow{\mu \succ T} \sum_{n=0}^{\infty}(-1)^n e^{-n\beta(\mu+p_0)} \qquad (10b)$$

$p_0$ correspond to the energy of the real particles, given by

$$p_0^2 = p^2 + m^2 \qquad (11a)$$

where, $p$ is the magnitude of the three momentum of the particles. In the presence of the constant background magnetic field, it reads out to be[19]

$$p_0^2 = p^2 + m^2 + eB(2n+1) \qquad (11b)$$

where $n$ corresponds to the Landau Levels. We ignore the polarization effects for the moment. The electromagnetic properties of neutrinos can be derived from the radiative decay of neutrino $\nu \to \nu\gamma$ or the Plasmon decay given as $\gamma \to \gamma\bar{\nu}$. These processes can only take place through the higher order processes. The most general form of the decay rate of Dirac neutrino can be written in terms of its form factors [5,6] as

$$\Gamma_\mu = \left[ F_1 \bar{g}_{\mu\nu} \gamma^\nu + F_2 u_\mu + iF_3 \left( \gamma_\mu u_\nu - \gamma_\nu u_\mu \right) q^\nu + iF_4 \varepsilon_{\mu\nu\alpha\beta} \gamma^\nu q^\alpha u^\beta \right] L \qquad (12)$$

in usual notation with

$$F_1 = T_T + \frac{\omega}{Q^2}(T_L - T_T) \qquad (13a)$$

$$F_2 = \frac{1}{V^2}(T_L - T_T) \qquad (13b)$$

$$iF_3 = -\frac{\omega}{Q^2}(T_L - T_T) \qquad (13c)$$

$$F_4 = \frac{T_P}{Q} \qquad (13d)$$

where, $L$, $T$ and $P$ correspond to the Longitudinal, Transverse and the Polarization components, respectively. $F_1$ here is in the form of standard charge radius, $F_2$ yields an additional contribution to the charge radius. $F_3$ and $F_4$ correspond to the electric and magnetic form factors of neutrinos, respectively. It can also be seen that

$$F_3(q^2 = 0) = D_E = -\frac{i\omega}{q^2}(T_L - T_T) \qquad (14a)$$

and
$$F_4(q^2 = 0) = D_M = -\frac{T_P}{2q} \tag{14b}$$

Where, $q$ on the right-hand side of the equation corresponds to the magnitude of three-momentum vector $q$, and $D_E$ and $D_M$ are the electric and magnetic dipole moment of Dirac neutrino. Considering

$$T_{\mu\nu} = T_T R_{\mu\nu} + T_L Q_{\mu\nu} + T_P P_{\mu\nu} \tag{15}$$

such that

$$R_{\mu\nu} \equiv \bar{g}_{\mu\nu} - Q_{\mu\nu} \tag{16a}$$

$$Q_{\mu\nu} \equiv \frac{\bar{u}_\mu \bar{u}_\nu}{\bar{u}} \tag{16b}$$

$$P_{\mu\nu} \equiv \frac{1}{|q|} \varepsilon_{\mu\nu\alpha\beta} q^\alpha u^\beta \tag{16c}$$

with,

$$\bar{g}_{\mu\nu} = g_{\mu\nu} - \frac{q_\mu q_\nu}{q^2} \tag{17a}$$

$$\bar{u}_\mu = g_{\mu\nu} u^\nu \tag{17b}$$

$$u_\mu = (1,0,0,0) \tag{17c}$$

for $\omega = q \cdot u$ and $q = (\omega^2 - |q|^2)^{1/2}$ and the form factors can be written as

$$T_T = \frac{eg^2}{2M^2}\left(\alpha - \frac{\beta}{u^2}\right) \tag{18a}$$

$$T_L = \frac{eg^2}{M^2} \frac{\beta}{u^2} \tag{18b}$$

$$T_P = -\frac{eg^2}{M^2}|q|\kappa \tag{18c}$$

The integrals $\alpha$, $\beta$ and $\kappa$ are given by

$$\alpha = \int \frac{d^3 p}{(2\pi)^3 2E}(n_F^+ + n_F^-)\left[\frac{2m^2 - 2p \cdot q}{q^2 + 2p \cdot q} + (q \leftrightarrow -q)\right] \tag{19a}$$

$$\beta = \int \frac{d^3 p}{(2\pi)^3 2E}(n_F^+ + n_F^-)\left[\frac{2(p \cdot u)^2 - 2(p \cdot u)(q \cdot u) - p \cdot q}{q^2 - 2p \cdot q} + (q \leftrightarrow -q)\right] \tag{19b}$$

$$\kappa = -\int \frac{d^3 p}{(2\pi)^3 2E}(n_F^+ - n_F^-)\left[\frac{1}{q^2 + 2p \cdot q} + (q \leftrightarrow -q)\right] \tag{19c}$$

Eq. (19) shows that the first two equations (19a) and (19b) give nonzero values of $\alpha$ and $\beta$ at any values of particle-antiparticle densities, whereas $\kappa$ in Eq. (19c) vanishes at low densities and is extremely small in CP- symmetric background. We have evaluated these integrals at $T \geq \mu$ in Ref.[6], giving

$$\alpha \cong \frac{1}{\pi^2}\left[\frac{c(m_\ell\beta,\mu)}{\beta^2}+\frac{m}{\beta}a(m_\ell\beta,\mu)-\frac{m^2}{2}b(m_\ell\beta,\mu)-\frac{m^4\beta^2}{8}h(m_\ell\beta,\mu)\right],$$

$$\beta \cong \frac{1}{\pi^2}\left[\left(1+\frac{3}{8}\ln\frac{1-v}{1+v}\right)\frac{c(m_\ell\beta,\mu)}{\beta^2}+\frac{m}{\beta}a(m_\ell\beta,\mu)-\frac{m^2}{2}b(m_\ell\beta,\mu)-\frac{m^4\beta^2}{8}h(m_\ell\beta,\mu)\right], \quad (20)$$

$$\kappa \cong \frac{1}{\pi^2}\left[b(m_\ell\beta,\mu)+m_\ell^2 h(m_\ell\beta,\mu)\right].$$

for negligible antiparticles background. However, in CP- symmetric background, we can write Masood's abc functions in Eqs.(20) as

$$a(m\beta,\pm\mu) = \ln\left(1+e^{-\beta(m\pm\mu)}\right), \quad (21a)$$

$$b(m\beta,\pm\mu) = \sum_{n=1}^{\infty}(-1)^n e^{\mp\beta\mu} Ei(-nm\beta), \quad (21b)$$

$$c(m\beta,\pm\mu) = \sum_{n=1}^{\infty}(-1)^n \frac{e^{-n\beta(m\pm\mu)}}{n^2}. \quad (21c)$$

$$h(m\beta,\pm\mu) \cong \sum_{n=1}^{\infty}(-1)^{n+1}\left(\frac{(n\beta)^2}{2}Ei(-nm_\ell\beta)+n\beta\frac{e^{-n\beta m_\ell}}{m_\ell}\right) \quad (21d)$$

Where $\pm\mu$ to the chemical potential of electron (positron).

$\kappa \to 0$ for $\mu \to 0$ as thermal contribution is extremely small. With nonzero chemical potential it depends on the density contribution. At large T, the chemical potential is negligible and we get the same background contribution for particles and antiparticles. Therefore in CP symmetric background Masood's abc functions (we call them $a_i$ functions for simplicity) for particles and antiparticles are evaluated in the same way. We define

$$a_{avg}(m\beta,\mu) = \frac{1}{2}[a(m\beta,\mu)+a(m\beta,-\mu)] \quad (22a)$$

$$a_{net}(m\beta,\mu) = \frac{1}{2}[a(m\beta,\mu)-a(m\beta,-\mu)] \quad (22b)$$

such that, in the limit $T \ll \mu$

$$a_{net}(m_\ell\beta,\mu) \to 0$$

or in other words the function $\kappa \to 0$, indicating the vanishing contributions to the magnetic moment of neutrino from the tadpole diagram corresponding to the polarization of the medium. In reference to Eq.(18c), $\kappa \to 0$, shows that the polarization factor of the form factors vanishes at extremely high temperatures. The calculational detail of these $a_i$ functions can be found in Ref.[5]. These functions always appear in this scheme of calculations. In the limit $T \geq \mu$,

Masood's functions read out as

$$a_{avg}(m_\ell, \beta, \mu) \simeq \frac{1}{2} \ln\left[\left(1 + e^{-\beta(m_\ell - \mu)}\right)\left(1 + e^{-\beta(m_\ell + \mu)}\right)\right] \quad (23a)$$

$$b_{avg}(m_\ell, \beta, \mu) \cong \sum_{n=1}^{\infty} (-1)^n \cosh(n\beta\mu) \, \mathrm{Ei}(-nm_\ell \beta) \quad (23b)$$

$$c_{avg}(m_\ell, \beta, \mu) \cong \sum_{n=1}^{\infty} (-1)^n \cosh(n\beta\mu) \frac{e^{-n\beta m_\ell}}{n^2} \quad (23c)$$

$$h_{avg}(m_\ell, \beta, \mu) \cong \sum_{n=1}^{\infty} (-1)^{n+1} \left(\frac{(n\beta)^2}{2} \mathrm{Ei}(-nm_\ell \beta) + n\beta \frac{e^{-n\beta m_\ell}}{m_\ell}\right) \cosh(n\beta\mu) \quad (23d)$$

whereas,

$$b_{avg}(m_\ell, \beta, \mu) \to 0 \quad (24a)$$

$$h_{avg}(m_\ell, \beta, \mu) \to 0 \quad (24b)$$

In the high density regime, at $\mu \leq T$ however, these functions are evaluated as the corresponding high density functions as $a_d$'s as

$$a_d(m_\ell \beta, \mu) \simeq \mu - m_\ell - \frac{1}{\beta} \sum_{r=1}^{\infty} \sum_{n=1}^{\infty} \frac{(-1)^n}{(n\beta\mu)^{1-r}} e^{-n\beta m_\ell} \cosh(n\beta\mu)$$

$$b_d(m_\ell \beta, \mu) \simeq \ln \frac{\mu}{m_\ell}$$

$$c_d(m_\ell \beta, \mu) \simeq \frac{\mu^2 - m_\ell^2}{2} - \mu^2 \sum_{r=1}^{\infty} (-1)^r \sum_{n=1}^{\infty} \frac{(-1)^n}{(n\beta\mu)^{2-r}} e^{-n\beta m_\ell} \cosh(n\beta\mu) \quad (25)$$

$$h_d(m_\ell \beta, \mu) \simeq \frac{\mu^{-4} - m_\ell^{-4}}{2} - \sum_{r=1}^{\infty} \frac{(-1)^r}{\mu^r} \sum_{n=1}^{\infty} \frac{(-1)^n}{(n\beta)^{4+r}} e^{-n\beta m_\ell} \cosh(n\beta\mu)$$

And the corresponding net functions can be calculated easily. However,

Integrals of Eqs.(12) help to understand the nature of the background contributions. At high temperatures, the background contributions due to fermion and anti-fermion distributions are approximately equal and hence they cancel each other such that

$$\kappa \to 0 \quad \text{as} \quad \mu \to 0 \quad (27a)$$

At high densities, however, particle and antiparticle contribution need not to be the same all the time which implies that

$$\kappa \neq 0$$

In the CP symmetric dense background, the chemical potential of the fermions is equal and opposite to the chemical potential of the anti-fermions. This is the situation where again the contribution due to fermions background would not be the same as the anti-fermion contribution because of the different sign of the chemical potential $\mu$ so Eq.(17b) holds true and we get,

$$\kappa \cong -\frac{1}{8\pi^2} \ln \frac{\mu}{m_\ell}$$

The magnetic moment of neutrino can simply be obtained from the bubble diagram of the Plasmon decay in the MSM. This contribution can be written as

$$\Lambda_0(p_1, p_2) = -\frac{g^2}{m_W^2} \int \frac{d^4k}{(2\pi)^4} \gamma_\alpha L(p_2 - k + m_\ell)\gamma_\mu(p_1 - k + m_\ell)\gamma^\alpha L \qquad (29)$$
$$\times \frac{2\pi i \delta[(p_2 - k)^2 - m_\ell^2]}{(p_2 - k)^2 - m_\ell^2} n_F(p_2 - k) + (p_1 ? p_2)$$

inthe standard notation of particle physics. We approximate the $W$ propagator as $1/m_W^2$ where $m_W$ is the mass of $W$ boson. The magnetic moment of neutrino contribution from this diagram is always nonzero for massive neutrinos and can be expressed in terms of neutrino mass as [6]

$$\alpha_\nu^\beta \approx \frac{G_F m_\ell}{4\pi^2} m_{\nu_\ell}\left[b(m\beta, \mu) + \frac{4}{M^2}\left(m\, a(m\beta, \mu) - c(m\beta, \mu)\right)\right]\mu_B \qquad (30)$$

with $\mu_B$, the Bohr Magneton. Eq.(30) gives the total background contribution at $(T > \mu)$ when $T$ is sufficiently greater than the chemical potential; a significant contribution comes from the last term only. So the approximate value of the magnetic moment is

$$a_{\nu_\ell} \approx \frac{T^2 G_F m_\ell m_{\nu_\ell}}{48 M^2} \mu_B \qquad (31)$$

where

$$c \to -\frac{\pi^2}{12} \quad \text{for } T \gg m_e \qquad (32)$$

Figure 3 gives a plot of magnetic moment as a function of temperature.

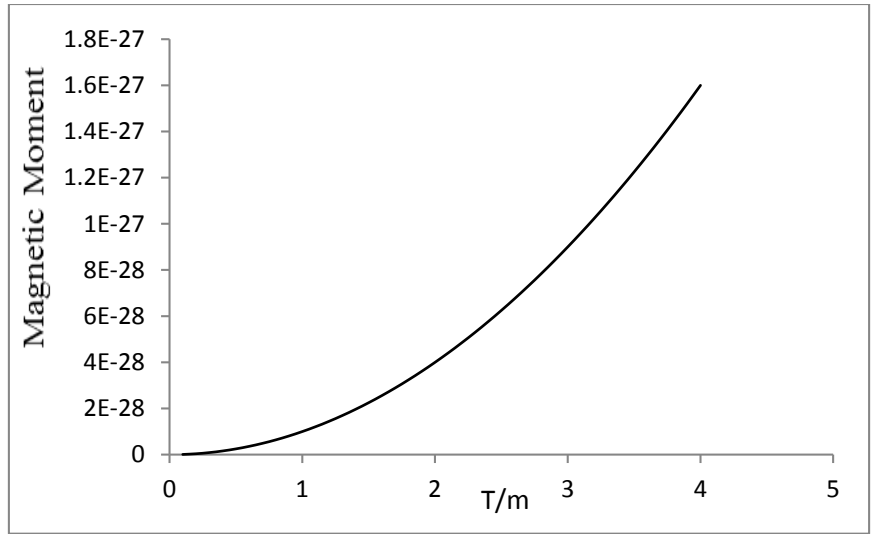

*Figure 3: Thermal correction to the magnetic moment of $\nu_e$ as a function of temperature.*

In the limit $\mu > T$, the magnetic moment contribution is coming from the tadpole diagram. Comparing Eq.(9b) with that of (10b), one can easily see the difference of behavior between the particle and antiparticle distribution functions. So, we have to evaluate the integral $\kappa$ to find the contribution from the tadpole diagram also. CP symmetry breaks at low temperature (comparable to electron mass). However, we are not considering the cold system so the statistical distribution function is still included in the calculations to incorporate the particle background effects. We first evaluate Eq.(29) to find the contribution of bubble diagram and $\kappa$ is evaluated to find the contribution of the tadpole diagram. Both of these contributions are added up together to give the statistical background contribution to the magnetic moment of Dirac neutrino of MESM in the high density regime as.

$$D_M^{stat} = \frac{eg^2}{2M^2}\kappa + \alpha_{\nu_\ell}^{stat}$$

$$D_E^{stat} = -\frac{i\omega}{q^2}\frac{eg}{2M^2}(3\beta - \alpha)$$

where as,

$$a_\nu^{stat} \cong \frac{G_F m_\ell}{8\pi^2} m_{\nu_\ell} \ln\frac{\mu}{m_\ell}$$

Eq. (34) indicates that the magnetic dipole moment of neutrino does not change much with the chemical potential $\mu$ because $\ln(\mu/m)$ is a slowly varying function of $\mu$. A plot of Eq. (21) is given in Figure 4.

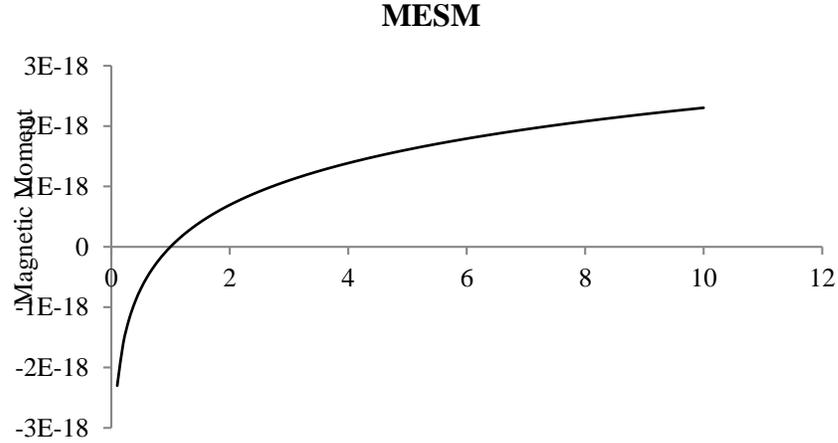

*Figure 4: A plot of magnetic moment as a function of chemical potential in units of electron mass in MESM*

and giving the magnetic moment in the units of the Bohr Magneton as,

$$D_M^{stat} \approx \left(\frac{G_F m_\ell}{8\pi^2} - \frac{m_\ell g_Z^2}{16M^2}\right) m_{\nu_\ell} \ln\frac{\mu}{m_\ell}$$

$M$ is the mass of $Z$ and $g_Z$ is the coupling constant of $Z \to \ell\bar{\ell}$ vertex. The relative sign

difference between the bubble diagrams indicates the decrease in the background contribution. The two terms in bracket cannot cancel each other because they still have the same type of contribution and the difference appears as a constant factor only. This factor is comparable to each other so provides a strong decreasing factor. We can safely say that this contribution is negligible for large chemical potentials. Eq.(34) shows that the magnetic moment at $\mu_e \sim 10 m_e$ is around $10^{-19}$ Bohr Magneton, which is even smaller than the vacuum value of the magnetic moment of neutrino which is around $10^{-18}$ Bohr Magneton [1,4]. In this situation, it seems as if the nonzero neutrino mass does not help to get the desired value of the magnetic moment of neutrino, even with the background correction.

In the extended standard models like HDM, the individual lepton no. is not conserve so the FTD corrections to the magnetic moment of neutrino could be different because we have to add new individual lepton number violating diagrams. The additional background FTD corrections to the magnetic moment of neutrino come out to be [6, 10]

$$a_{\nu_i} = \frac{f_{\ell\ell'} f_{\ell\ell'}}{16\pi^2 m_h^2} m_\ell^2 \left[ \sum_{n=1}^{\infty} (-1)^n e^{-n\beta\mu} Ei(nm\beta) \right] \mu_B \square -3 \times 10^{-19} \left( \frac{m_{\nu_i}}{eV} \right) \mu_B \quad (36)$$

with

$$f_{e\mu} h_{e\mu} / m_h^2 \to 2.8 \times 10^{-6} \text{ GeV}^{-2}$$
$$f_{e\tau} h_{e\tau} / m_h^2 \to 2.8 \times 10^{-6} \text{ GeV}^{-2}$$

Thermal corrections in Eq. (36) are not only too small but negative which actually give a further decrease. $m_h$ is the mass of the charged Higgs in the new Higgs doublet in this model. Eq.(29) leads to

$$a_{\nu_\ell}^{stat} \approx \sum_{\ell'=e,\mu,\tau} \frac{f_{\ell\ell'} h_{\ell\ell'}}{16\pi^2 m_h^2} m_\ell^2 \ln\frac{\mu}{m_\ell} \mu_B$$

at $\mu \succ T$. Figure 5 shows a plot of magnetic dipole moment as a function of

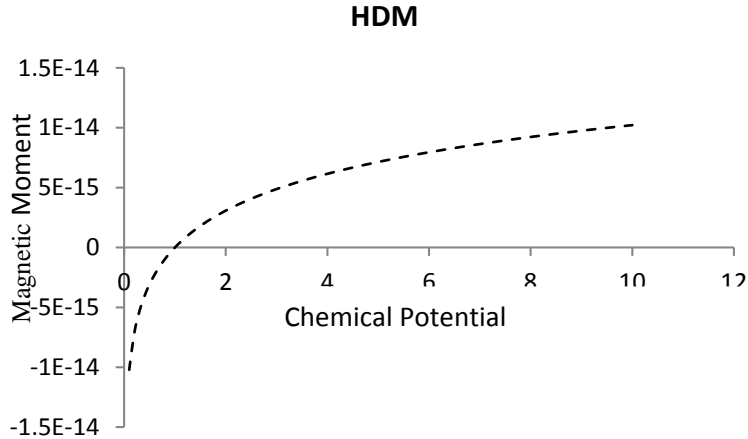

*Figure 5: A plot of magnetic moment as a function of chemical potential in units of electron mass in HDM*

### 3.1 Majorana Neutrino

The magnetic moment of neutrino changes strongly with chemical potential. It is well-known that the properties of Majorana neutrinos are different from that of Dirac neutrino. So, it cannot have the dipole moment as we describe for Dirac neutrino. However, it can have a transition magnetic dipole moment. Similarly the neutrinos of the SUSY models canalso have the transition magnetic moment. Riotto [21] has claimed that a large class of models where charged scalar boson couple to leptons can provide the magnetic moment of electron type neutrino as large as $10^{-12}$ Bohr Magneton, which can play a relevant role in different astrophysical phenomenon. However, these extended standard models are constructed to get the required value of magnetic moment but the direct testing of the model is not easy. However, effective mass of Weyl neutrino could be considered a natural solution with the effective mass. Magnetic moment of different flavor neutrinos will then depend on the chemical potential of the corresponding charged fermions and their corresponding statistical conditions.

Majorana neutrinos undergo seesaw mechanism and it can acquire transition magnetic moment as the lepton number conservation is always violated. Like all other models, form factors are induced through radiative corrections. For Majorana particles, radiative corrections are model dependent. A comparison with MESM is not meaningful in the presence of a larger particle sectors as well as different conservation rules of the models with Majorana neutrinos. Heavy neutrinos, seesaw mechanism and heavy scalars with supersymmetric partners [21-23] make situation much more different. So the calculations of form factors of neutrino is model dependent and has to be evaluated in different rages of temperature and chemical potential separately.

### 4. Results and Discussions

Magnetic moment is one of the electromagnetic properties of matter and is proportional to mass.

$$\mu_{\nu_i} = \frac{3 m_e G_F}{4\sqrt{2}\pi^2} m_{\nu_i} \cong 3 \times 10^{-18} \left(\frac{m_{\nu_i}}{\text{eV}}\right) \mu_B$$

Vacuum values of neutrino magnetic moments for different flavors of neutrinos and is given by

| $\ell$ | Mass (MeV/c$^2$) | MESM ($\mu_B$) |
|---|---|---|
| $\nu_e$ | $< 2.2 \times 10^{-6}$ | $0.3 \times 10^{-19}$ |
| $\nu_\mu$ | $< 0.17$ | $0.3 \times 10^{-14}$ |
| $\nu_\tau$ | $< 18.20$ | $3.0 \times 10^{-12}$ |

There is no background correction for massless neutrinos even at high temperatures as temperature does not contribute to the effective mass of massless particles. The QED type radiative corrections are only important below the decoupling temperatures [24.25]. We therefore study the magnetic moment below $T \leq 4 m_e$. However, theeffective mass of neutrino is generated by the coupling of neutrino with the extremely dense background with constant magnetic field as is given in Eq. (1). Rate of change of mass also depends on the strength of the magnetic field and is given in Eq. (2). Figure 1 and Figure 2 show that the effective mass does not change significantly with the magnetic field until and unless this field is extremely large, really close to the threshold value, where it can go to infinity. However, the rate of change of mass (Figure 2) is even slowly varying as compared to the effective mass (Figure 1) itself. For Dirac type of mass of neutrino in the MESM the charged current contribution is nonzero at all temperatures and density and it is proportional to the mass of neutrino [4], as expected. This contribution comes from the bubble diagram and corresponds to charged current interaction. Tadpole diagram of neutral current interaction does not contribute as particle-antiparticle behavior is exactly same at finite temperatures. However, the high density contributions (with extremely large values of $\mu$) change the situation. Chemical potential of particles and antiparticles may not be equal, even in magnitude. So the induced magnetic dipole moment of Dirac neutrino picks up a nonzero contribution from a tadpole diagram also.

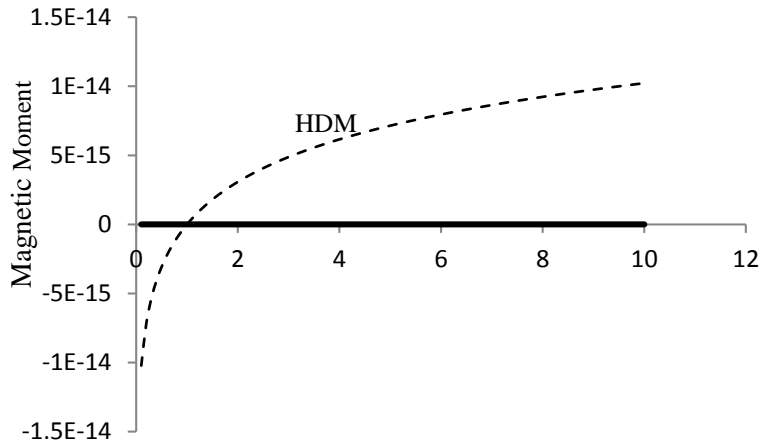

*Figure 6: Chemical potential corrections to the magnetic moment of electron type neutrino for MESM(solid line) and HDM(broken line).*

It is clear from the above results that the temperature corrections are really significant in the MESM only HDM gives a negligibly small negative contribution so its effect is the decrease instead of increased magnetic moment. However, the HDM pushes the magnetic moment limit faster to required values at chemical potential as compared to the MESM. A comparison of magnetic moment in HDM (dotted Line) and MESM(solid Line) is given in Figure 6.

The above analysis shows that the magnetic dipole moment of electron type neutrinos does not depend significantly on temperature so thermal corrections do not give enough corrections to touch the cosmological limit [6]. However, chemical potential contribution to the magnetic moment is non-ignorable, even when chemical potential is a little more than the electron mass. Figure 6 shows that the density effect (or chemical potential dependence) is more significant in the extended Higgs sector with individual lepton number violation. It gives enough motivation to study the statistical corrections to the magnetic moment of neutrino in Supersymmetric models to look for an appropriate solution to some of the cosmological issues. This result is more relevant for the superdense stars such as neutron stars [26] or magnetars where magnetic field is extremely high and apparently small contribution is not insignificant. However, the magnetic moment of neutrino and its thermal corrections are negligible in the early universe and does not give significant contribution to lepton scatterings [27,28] in the early universe.

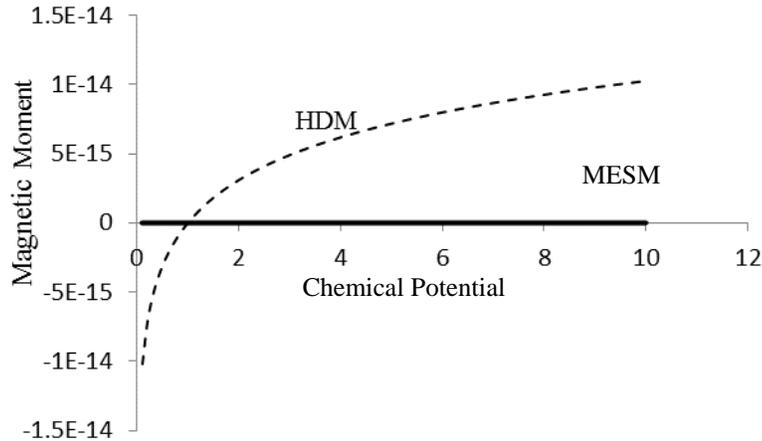

*Figure 6: Chemical potential corrections to the magnetic moment of electron type neutrino for MESM(solid line) and HDM(broken line).*

**References and Footnotes**

1) R.N.Mohapatra and P.B.Pal, Massive Neutrinos in Physics and Astrophysics, (1991) (World Scientific Publication), and references therein.
2) J.Schechter and W.F.Valle, Phys.Rev. D24, 1883 (1981).
3) Lepton dipole moments (ed B. Roberts, J. Marciano) world Scientific, (2010).
4) B.W. Lee and R. E. Shrock, Phys.Rev.D16, 1444(1977).
5) SaminaS.Masood, Phys.Rev.D48, 3250 (1993) and references therein.
6) SaminaS.Masood, Astroparticle Physics, 4, 189 (1995); **arXiv: hep-ph/0109042**
7) T.Kajita, Nucl.Phys.Proc.Suppl.77:123-132 (1999). http://arxiv.org/pdf/hep-ex/9810001v1.pdf
8) M.Koshiba, `Kamiokande and Superkamiokande' , Fifth School on Non-Accelerator Particle Astrophysics, Abdus Salam ICTP, Trieste, Italy (July 1998).
9) http://www-sk.icrr.u-tokyo.ac.jp/whatsnew/DOC-20110615/KEK110613english.pdf
10) K.S.Babu and V.S.Mathur, Phys.Lett. B196, 218(1987).
11) M.Fukujita and T.Yanagida, Phys.Rev.Lett.58, 1807(1987).
12) See for example: P.Landsman and Ch. G. Weert, Phys.Rep.145, 141 (1987) and references therein.
13) See for Example, K.Ahmed and SaminaSaleem (Masood) Phys.Rev. D35, 1861(1987).
14) K.Ahmed and SaminaSaleem (Masood), Phys.Rev. D35, 4020 (1987).
15) K.Ahmed and SaminaSaleem (Masood), Ann.Phys.164, 460(1991).
16) SaminaS.Masood, Phys.Rev.D44, 3943(1991).
17) SaminaS.Masood, Phys.Rev.D47, 648(1993).
18) Samina S. Masood and MahnazHaseeb, Astroparticle Physics 3 (4), 405-412 (1995).
19) A. Perez-Martinez, et al.; `Effective Magnetic Moment of Neutrinos in Strong Magnetic